\documentclass{article}
\usepackage{graphicx}
\usepackage[round]{natbib}
\usepackage{epsfig}
\title{
An objective change point analysis of \emph{landfalling} historical
Atlantic hurricane numbers }

\author{Stephen Jewson (RMS)\footnote{\emph{Correspondence email}: \texttt{stephen.jewson@rms.com}}\\
Jeremy Penzer (LSE)\\}

\begin{document}
\maketitle

\begin{abstract}
In previous work we have analysed the Atlantic basin hurricane
number time-series to identify decadal time-scale change points.
We now repeat the analysis but for US \emph{landfalling} hurricanes. The
results are very different.
\end{abstract}

\section{Introduction}

In previous work~\citep{e02a} we have attempted to identify change
points in the Atlantic hurricane number time-series. We used a
brute-force search through all possibilities to find the
change points defined by the global minimum in an out-of-sample
mean squared error (MSE) cost function. When we allowed gaps between change points as
short as 2 years we were unable to find the global minimum because
of the vast number of possible combinations of change points
relative to currently available computer power. However, when we increased
the minimum gap between change points to 10 years we were able to
find the global minimum of the cost function. This global minimum
corresponded to 4 change points, occurring in 1931/1932,
1947/1948, 1969/1970 and 1994/1995.
We also applied the method to intense storms only, and found change-points
in the years 1914/1915, 1947/1948, 1964/1965 and 1994/1995. The most recent two change-points we identify
in the intense series correspond exactly with the most recent two
change-points identified in the earlier work of
\citet{elsnerj00, elsnern04}, using a very different detection algorithm.

We now repeat our change-point analysis for US landfalling hurricanes
only. This is a much more difficult time series to work with,
since the number of landfalling hurricanes is much lower than the
total number of Atlantic hurricanes, and there are many years with
no landfalling hurricanes at all. We can thus anticipate that it
might be more difficult to identify change points in this
time-series. \citet{elsnern04} has previously analysed the same
time series using a Markov Chain Monte Carlo method. In section 5
we compare our results with those from this earlier study.

\section{Methods}

Our method is the same as that used in~\citet{e02a}, except that we now consider US landfalling hurricane numbers
rather than basin hurricane numbers.
We also only use a 10 year minimum gap between change points, whereas in~\citet{e02a} we performed a preliminary
study using a 2 year minimum gap.

\begin{figure}[!hb]
  \begin{center}
    \scalebox{1.2}{\includegraphics{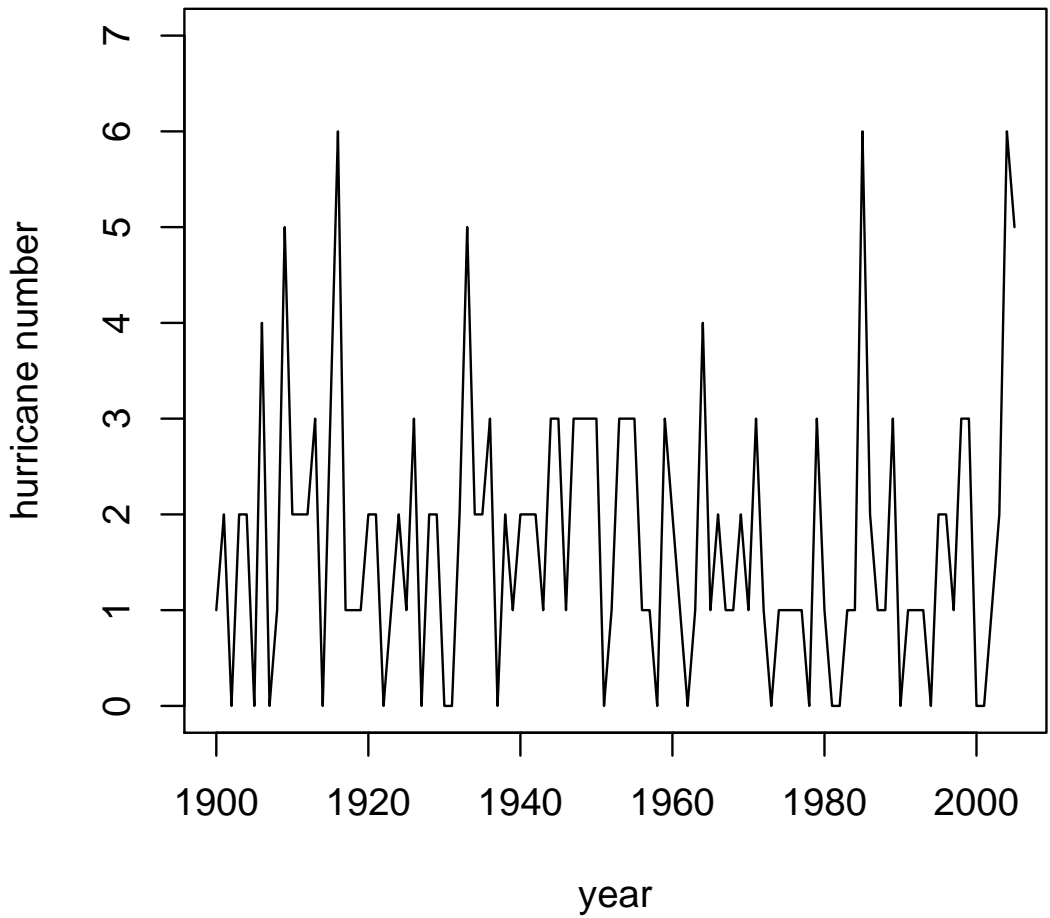}}
  \end{center}
    \caption{
US landfalling hurricane numbers for the period 1900 to 2005. }
     \label{f01}
\end{figure}

We create our landfalling hurricane number time series directly from the current version of the
HURDAT database~\citep{hurdat}, and define a landfalling hurricane as one which is a hurricane at the point of
landfall. We exclude hurricanes that weaken to non-hurricane status before landfall. The resulting
time series of hurricane numbers is shown in figure~\ref{f01}.

\section{Results}

Tables~\ref{10t01}, \ref{10t02} and~\ref{10t03} show the change
points detected in the landfalling hurricane number time-series,
the out-of-sample RMSE scores for these change points, and the number of
combinations tested to find them, respectively. Starting with the
RMSE scores in table~\ref{10t02}, we see a big difference between
these results and the results in~\citet{e02a}. In that study, the RMSE
reduced monotonically as we increased the number of levels, down
to a minimum at 5 levels, after which it started to increase
again. The minimum RMSE achieved was highly statistically significant
i.e. it is very unlikely that it could have occurred unless the hurricane
number time series has real temporal structure.
In the landfalling case, however, there is no
clear monotonic decrease. In fact the best two level model is slightly \emph{worse} than the
one level model.
The best of the models with more than 2 change-points beat the score for the one level model, but only very slightly.
We cannot, therefore, identify any change-points at all in
this time series, and a model which considers the landfalling rate to be
constant in time performs as well as any other.

We could, at this point, stop this study, since we haven't found
any statistical evidence for change points in the landfalling
hurricane time-series. We will, however, press on, since we have
strong physical reasons to think that change points do actually
exist. The total number of hurricanes shows clear change points,
and landfalling hurricanes are closely related, both physically
and statistically, to the total number of hurricanes. We now ask:
even though we can't detect any benefit from the modelling of
change points in the landfalling time-series, what
\emph{indications} are there that there might be change points in
this time series? If we \emph{had} to choose some change points,
what would they be? And do the change points we find show any
resemblance to those in the total number of hurricanes?

\begin{figure}[!hb]
  \begin{center}
    \scalebox{0.8}{\includegraphics{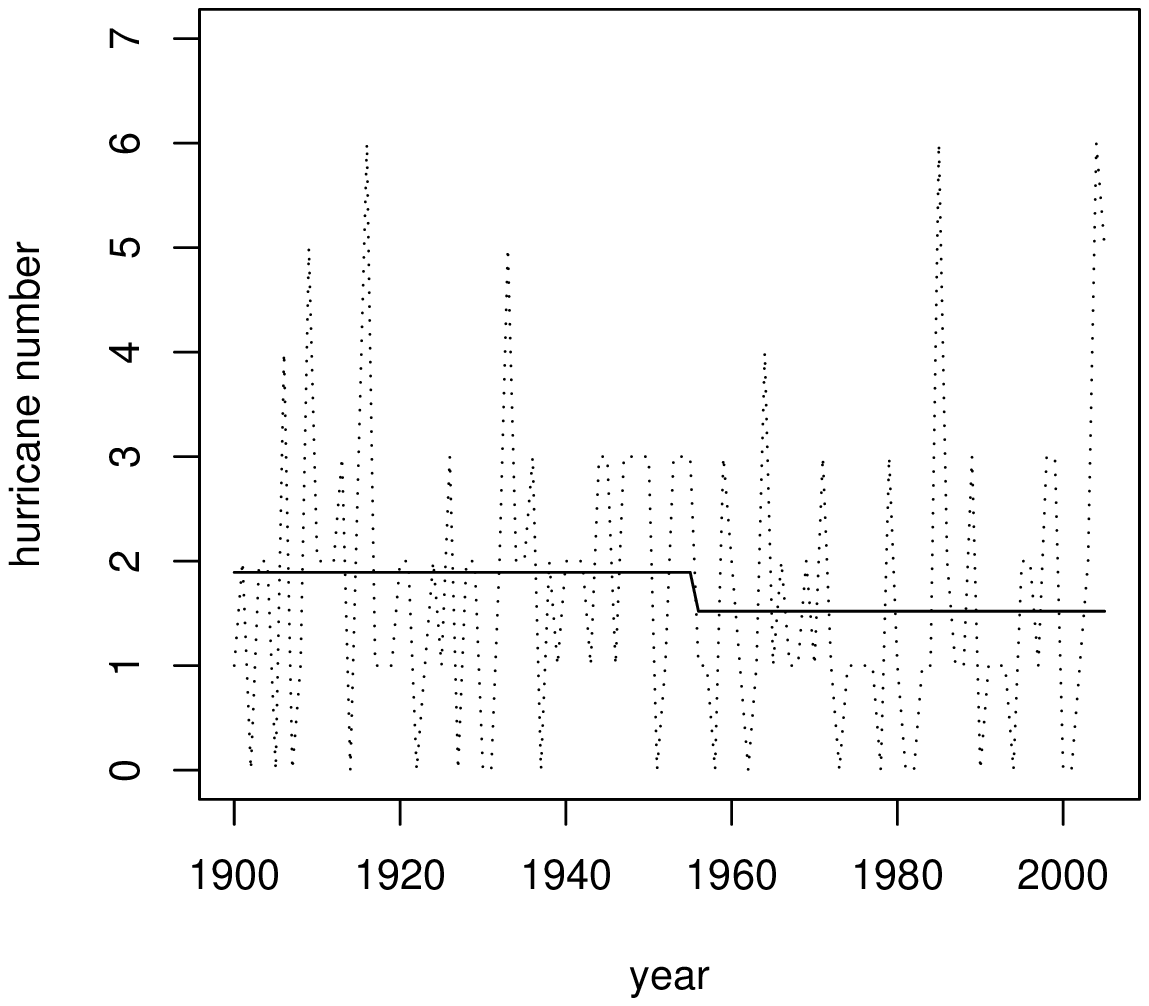}}
  \end{center}
    \caption{
The best 2 level model.
}
     \label{10f01}
  \begin{center}
    \scalebox{0.8}{\includegraphics{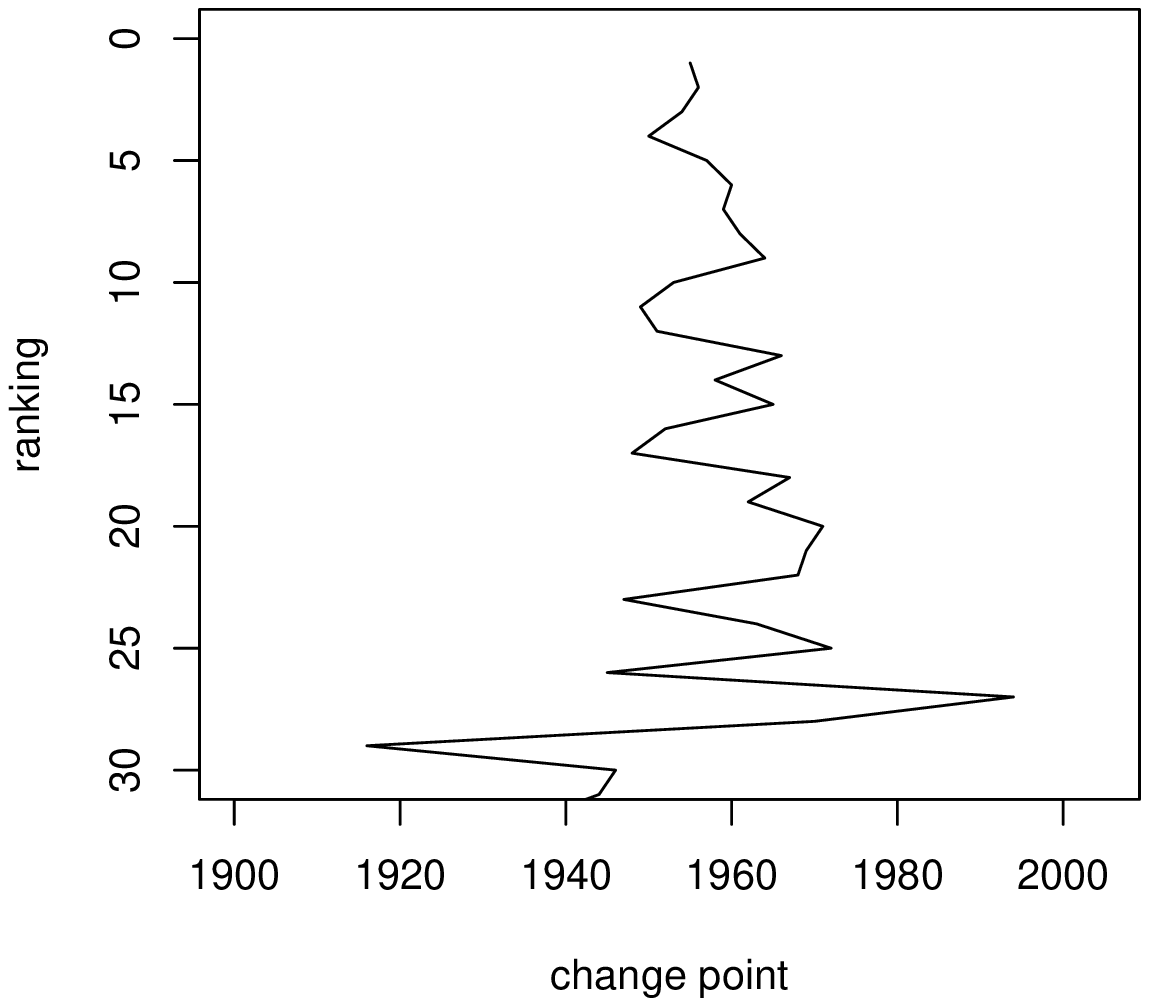}}
  \end{center}
    \caption{
The change points for the top 30 two level models considered.
}
     \label{10f02}
\end{figure}

Figure~\ref{10f01} and subsequent figures show the locations of
change points for the optimum models with 1 to 4 change points.
The one change point results show a reduction in the
number of hurricanes in 1956. The two change point results show a
brief reduction in activity in the 1970s and 1980s, which is
perhaps similar to the reduction in activity from 1970 to 1994 seen in the total number
results (see figure 9 in~\citet{e02a}). The three change point results
show a brief increase in the 1950s, which again is perhaps similar
to something seen in the total number results. The four change
points results show reductions in the 1920s and the 1960s-1970s.
Overall, however, we have to conclude that the change points we
identify don't really match closely with the change points for the
basin, although with the eye of a believer one can perhaps see
some similarities. Considering the stability of our optimal change
points, relative to the top 30 results in each case, we see that
the results are distinctly less stable than the equivalent results
for basin numbers.

\section{Intense landfalling hurricanes}

We now repeat our change-point analysis for intense landfalling hurricanes. In
the basin data the change points for the intense hurricanes are
more visually striking than those for the total number of
hurricanes, and so one might imagine that it might be possible to
detect change points in the intense number of landfalls even if it
is not possible to detect them in the total number of landfalls.

Figure~\ref{if01} shows the time-series of the number of intense landfalling hurricanes.
By eye, it is hard to see any marked changes in this time series.

Tables~\ref{it01} and~\ref{it02} show the change points
identified, with the scores. At least the score now does
decrease as we move from 0 change points (1 level) to 2 change
points (3 levels), suggesting that \emph{perhaps} there might be
two real change points in this data. However, as we saw in the
statistical tests in~\citet{e02a}, the presence of a minimum in this score
is actually no indication of statistical significance: what
matters is the value at the minimum. Running the same set of
statistical tests we ran in~\citet{e02a}, we find that the minimum
achieved for the intense landfalling data is not significant at all. Of our 100
random reorderings of the data, the mean of the best RMSE values
achieved was 0.822, which is actually \emph{lower} than the 0.840
achieved by the optimal RMSE value on the real data.

\section{Discussion}

We have performed an objective change point analysis on data for
the number of hurricanes making landfall annually on the US
coastline. We did not find any evidence for the existence of
change points in this time-series. The change points that our
algorithm does detect do not correspond very closely to the change
points that we have found previously in the total number of
hurricanes, and are not particularly stable. These results are consistent
with the earlier findings of \citet{elsnerj04}, who ran a very similar
analysis but using a very different algorithm based on Monte Carlo Markov
Chains rather than brute-force searching.

An interesting avenue of
future work would be to examine regionalization, and apply our
change point analysis to the Florida, East Coast and Gulf Coast landfalling
hurricane number time series separately. \citet{elsnerj04} found that the only detectable change points were
in the Florida time series.

In conclusion: the average number of landfalling hurricanes may change over time,
but there is little evidence for such changes in the landfalling hurricane data considered here.
This could either be because such changes don't exist, or, perhaps more likely, because there is not enough
data to distinguish the changes from the noise.


\section{Acknowledgements}

Thanks to Manuel Lonfa, Roman Binter and Shree Khare for interesting discussions on this topic.

\bibliography{arxiv}

\newpage

\newpage
\begin{table}[h!]
  \centering
\begin{tabular}{|c|c|c|c|c|c|c|}
 \hline
 1 & 2 & 3 & 4 & 5 & 6 & 7\\
 \hline
 model& cp1 & cp2 & cp3 & cp4 & cp5 & cp6\\
 \hline
1&&&&&&\\
2&  56&&&&&\\
3&  72&  85&&&&\\
4&  44&  56&  85&&&\\
5&  17&  32&  56&  85&&\\
6&  17&  32&  56&  72&  85&\\
7&  17&  32&  44&  56&  72&  85\\

 \hline
\end{tabular}
\caption{ The change points identified in the landfalling
hurricane number time series, versus the number of model levels. }
\label{10t01}
\end{table}

\begin{table}[h!]
  \centering
\begin{tabular}{|c|c|}
 \hline
 1 & 2\\
 \hline
 model& predictive RMSE\\
 \hline
 1&   1.412481    \\
 2&   1.413486    \\
 3&   1.399329    \\
 4&   1.401652    \\
 5&   1.396658    \\
 6&   1.392119    \\
 7&   1.398182    \\

 \hline
\end{tabular}
\caption{ The predictive RMSE scores for the different
models.}\label{10t02}
\end{table}
\begin{table}[h!]
  \centering
\begin{tabular}{|c|c|}
 \hline
 1 & 2\\
 \hline
 model& number of combinations tested\\
 \hline
1&  1.00E+00\\
2&  8.70E+01\\
3&  5.93E+03\\
4&  3.01E+05\\
5&  1.06E+07\\
6&  2.29E+08\\
7&  2.57E+09\\

 \hline
\end{tabular}
\caption{ The number of combinations tested for each
model.}\label{10t03}
\end{table}

\newpage

\newpage
\begin{figure}[!hb]
  \begin{center}
    \scalebox{0.8}{\includegraphics{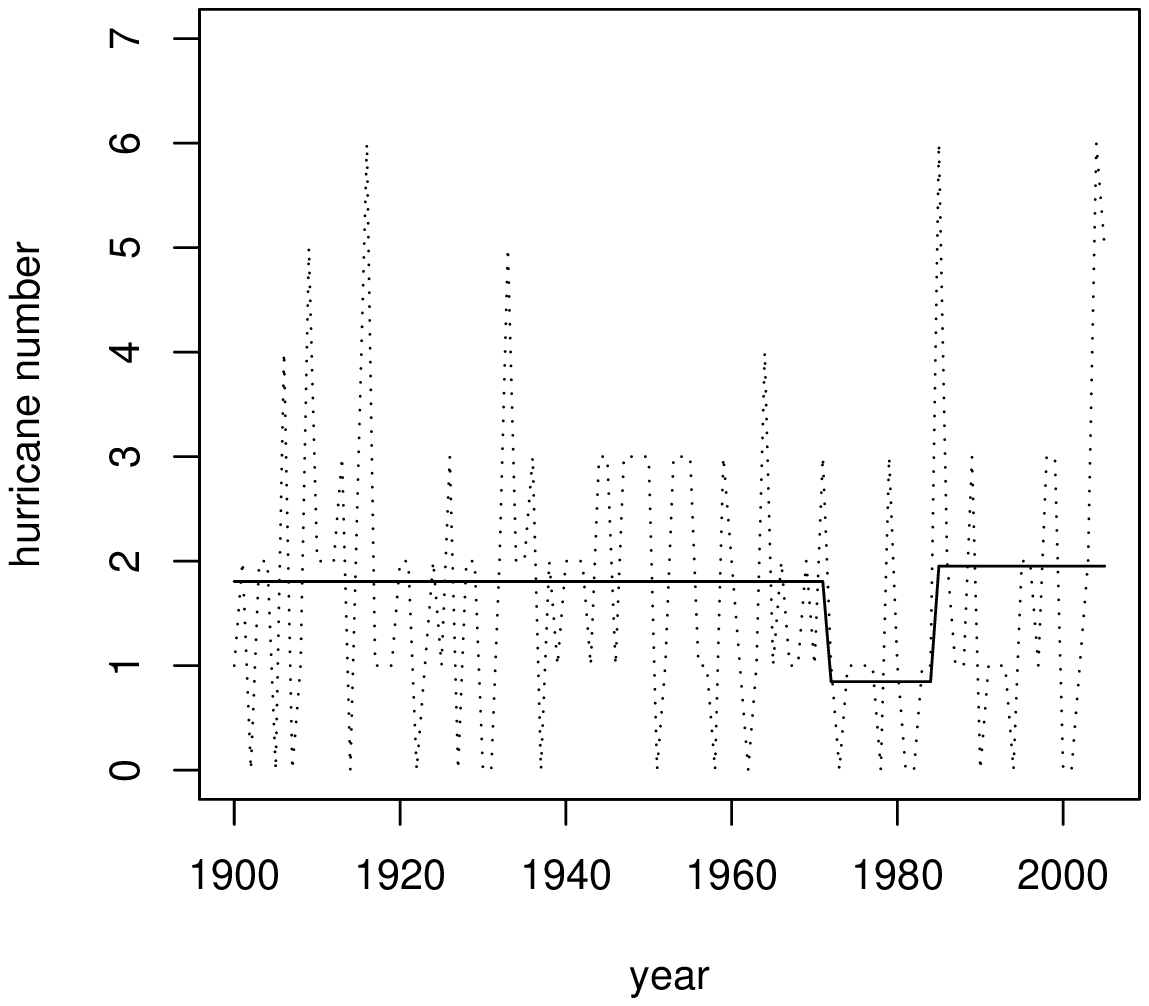}}
  \end{center}
    \caption{
The best 3 level model. }
     \label{10f03}
  \begin{center}
    \scalebox{0.8}{\includegraphics{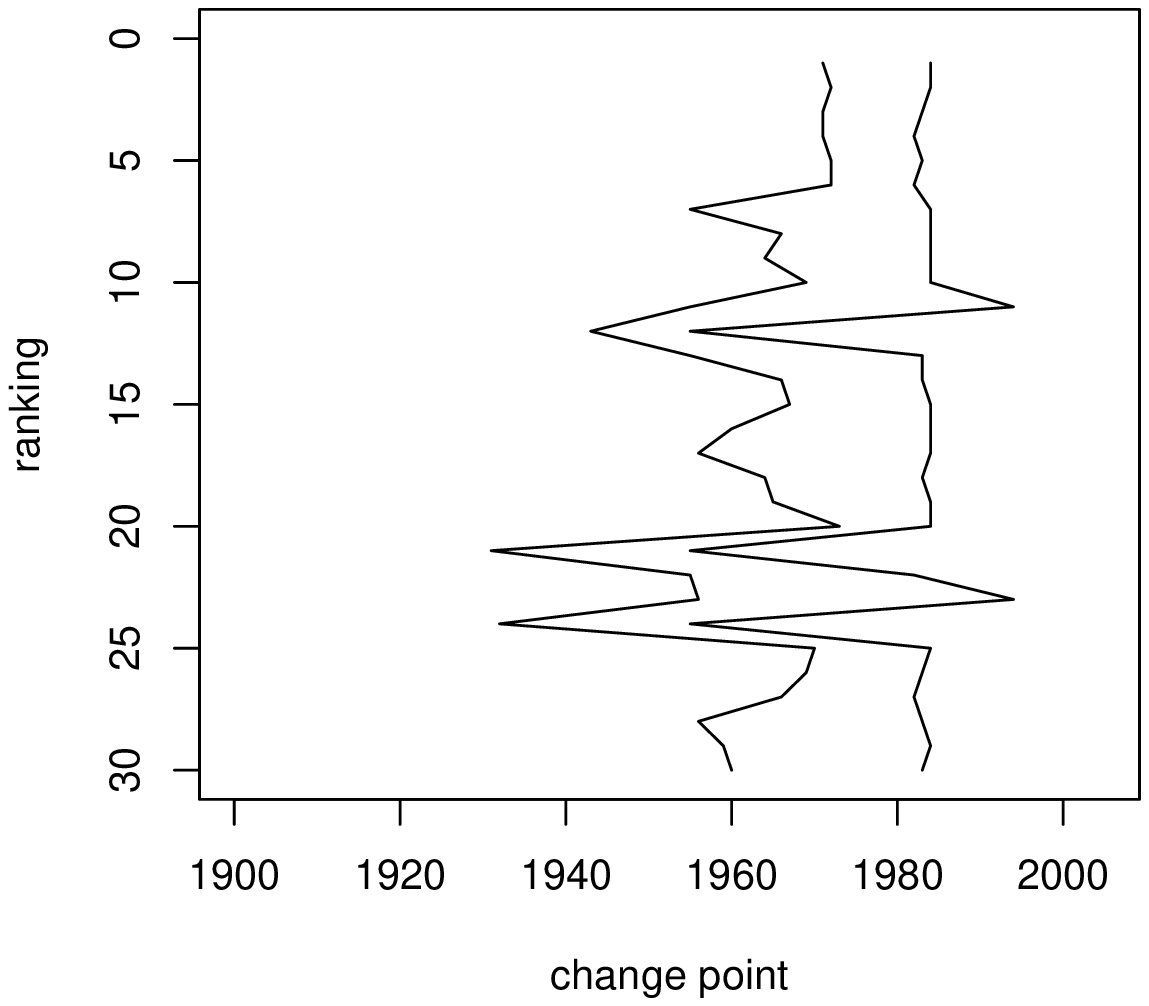}}
  \end{center}
    \caption{
The change points for the top 30 three level models. }
     \label{10f04}
\end{figure}

\newpage
\begin{figure}[!hb]
  \begin{center}
    \scalebox{0.8}{\includegraphics{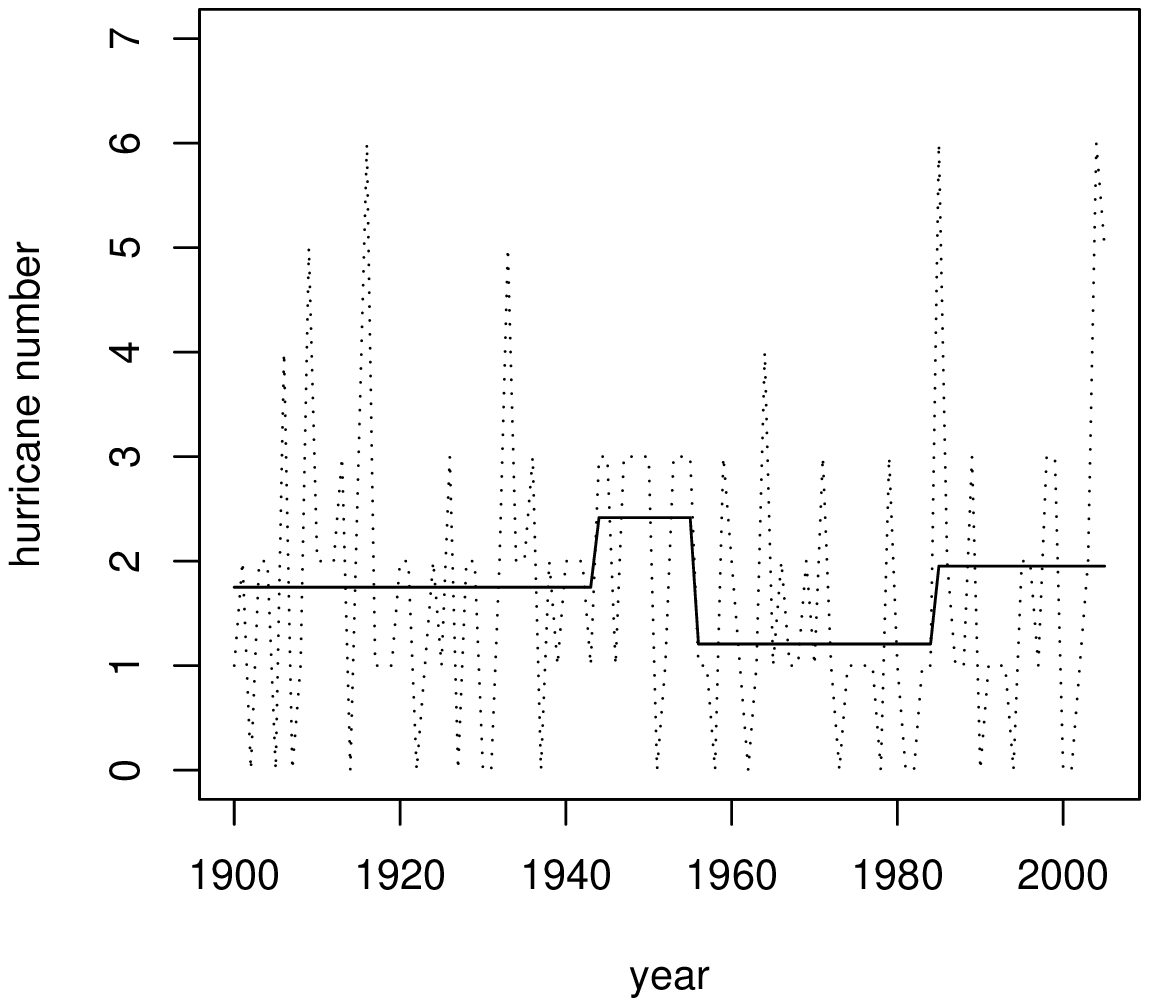}}
  \end{center}
    \caption{
The best 4 level model.
}
     \label{10f05}
  \begin{center}
    \scalebox{0.8}{\includegraphics{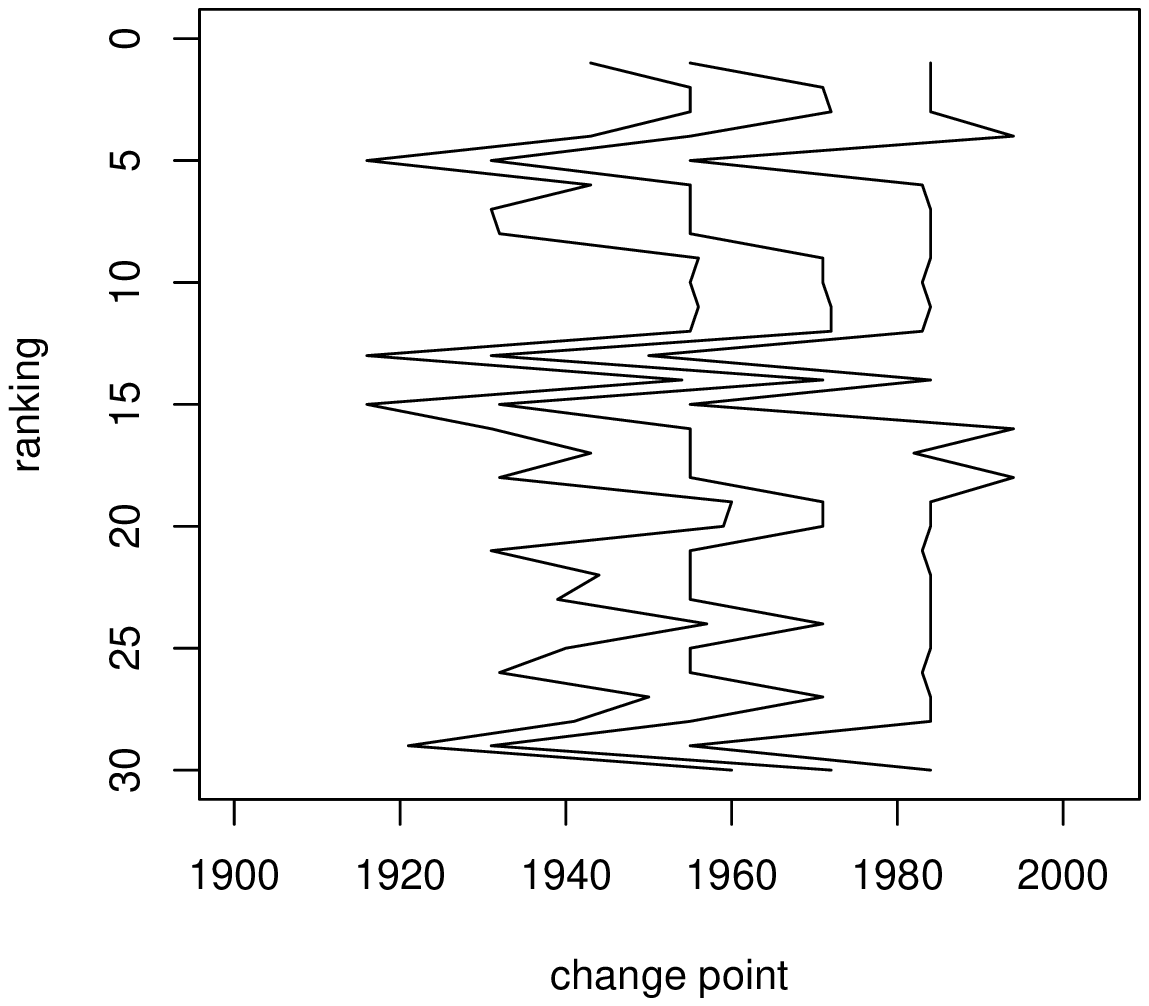}}
  \end{center}
    \caption{
The change points for the top 30 four level models.
}
     \label{10f06}
\end{figure}

\newpage
\begin{figure}[!hb]
  \begin{center}
    \scalebox{0.8}{\includegraphics{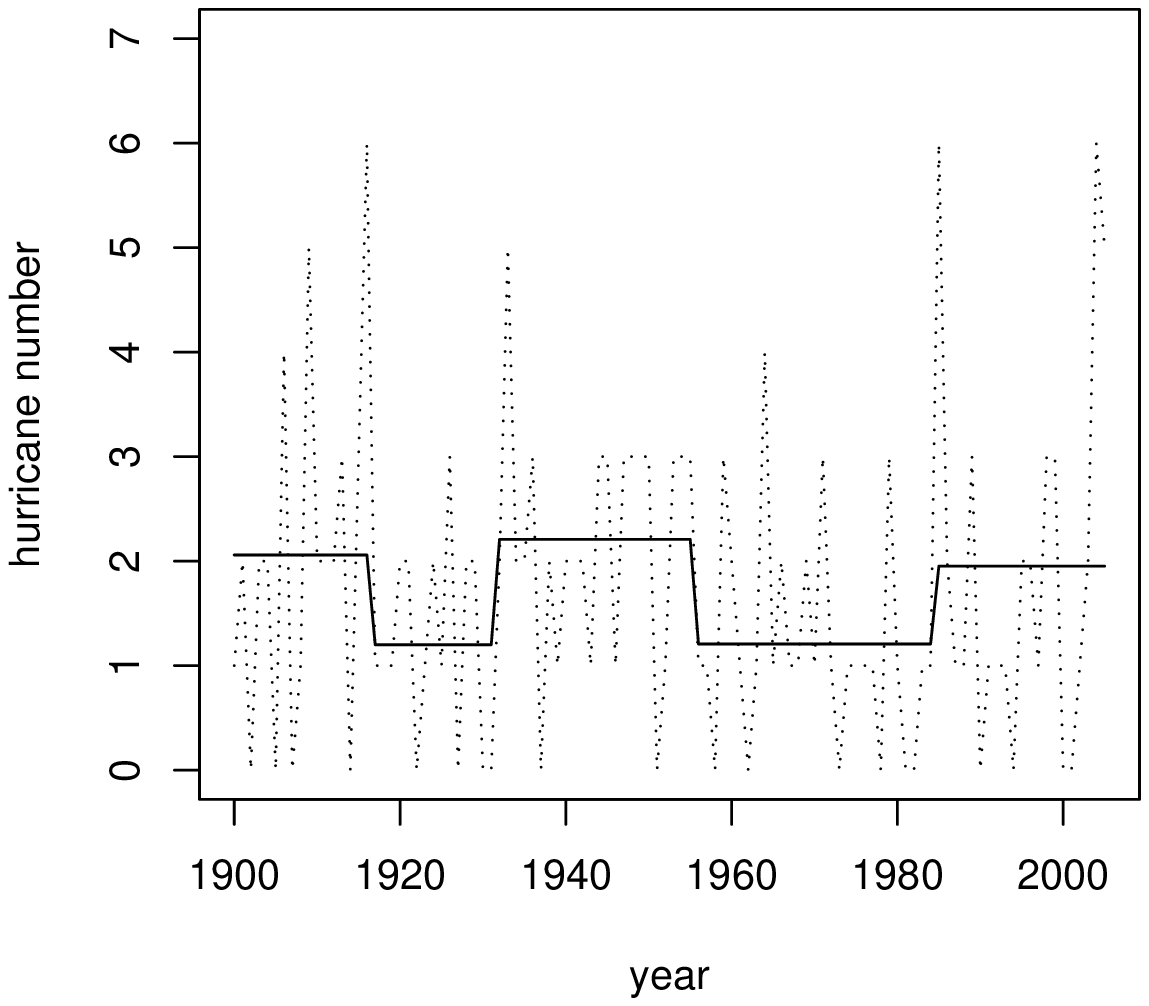}}
  \end{center}
    \caption{
The best 5 level model.
}
     \label{10f07}
  \begin{center}
    \scalebox{0.8}{\includegraphics{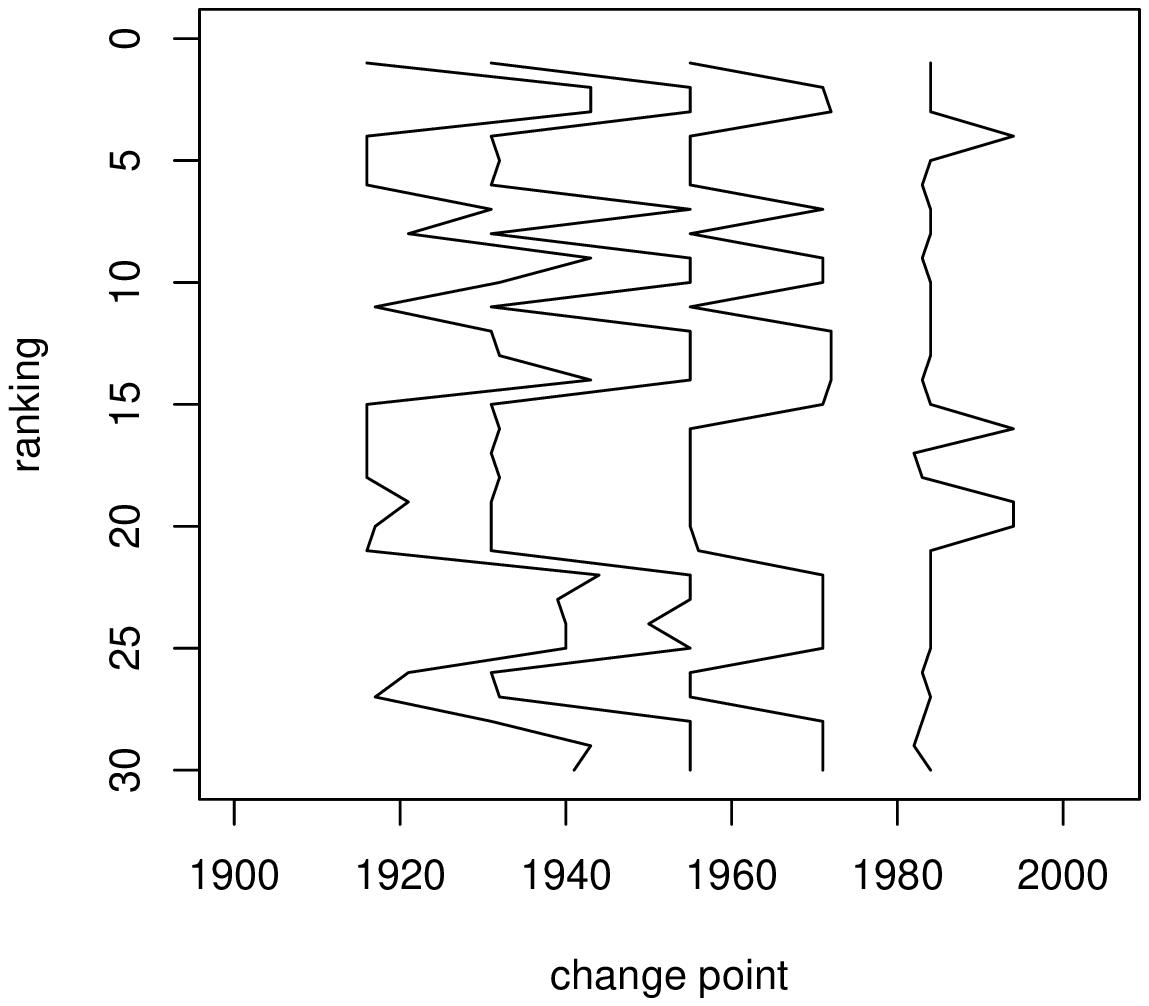}}
  \end{center}
    \caption{
The change points for the top 30 five level models.
}
     \label{10f08}
\end{figure}

\newpage
\begin{figure}[!hb]
  \begin{center}
    \scalebox{1.2}{\includegraphics{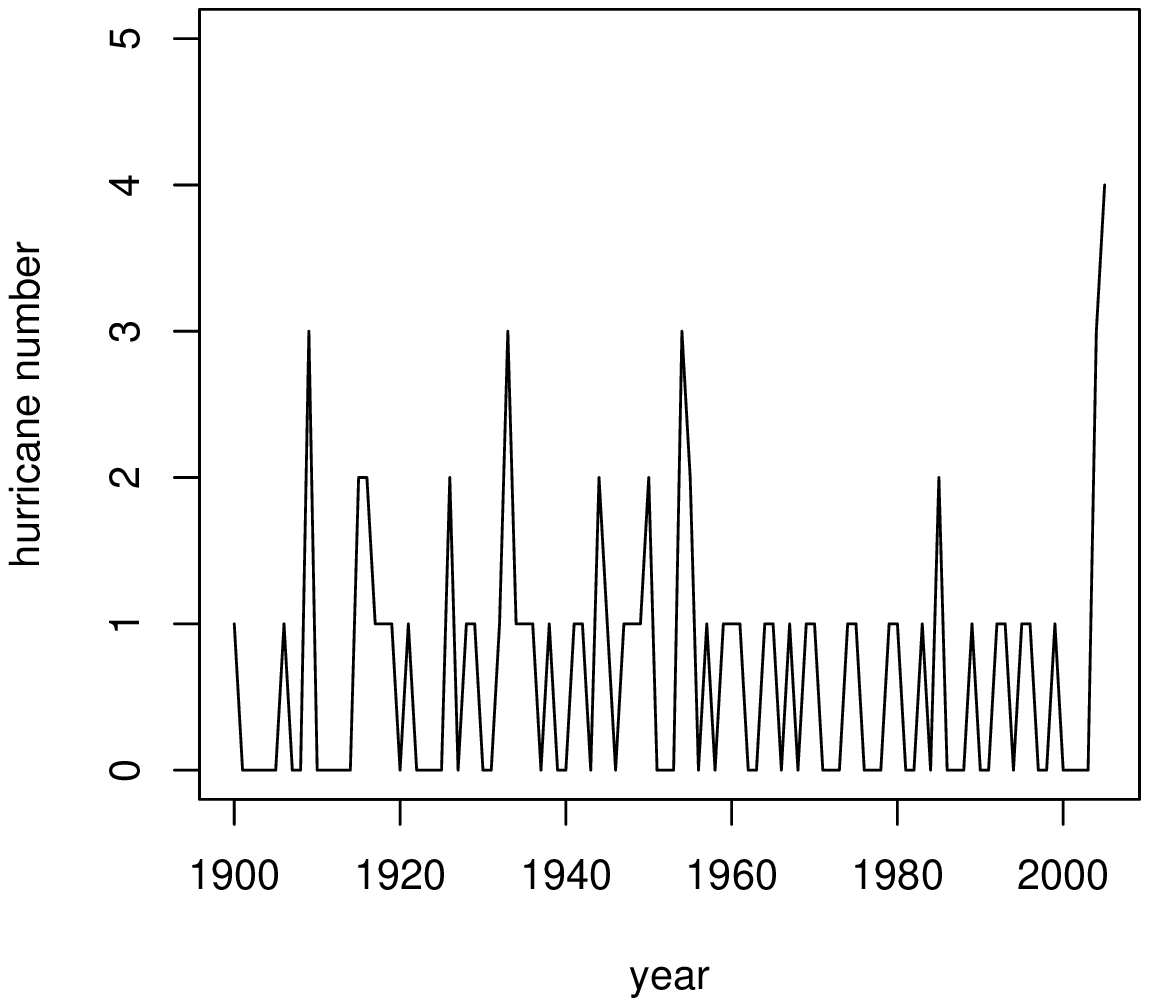}}
  \end{center}
    \caption{
US landfalling hurricane numbers for the period 1900 to 2005. }
     \label{if01}
\end{figure}

\newpage
\begin{table}[h!]
  \centering
\begin{tabular}{|c|c|c|c|c|c|c|}
 \hline
 1 & 2 & 3 & 4 & 5 & 6 & 7\\
 \hline
 model& cp1 & cp2 & cp3 & cp4 & cp5 & cp6\\
 \hline
1&&&&&&\\
2&  15&&&&&\\
3&  15&  56&&&&\\
4&  15&  62&  79&&&\\
5&  15&  56&  71&  83&&\\
6&  15&  32&  56&  71&  83&\\
7&  15&  34&  44&  56&  71&  83\\

 \hline
\end{tabular}
\caption{ The change points identified in the landfalling
\emph{intense} hurricane number time series, versus the number of
model levels. } \label{it01}
\end{table}

\begin{table}[h!]
  \centering
\begin{tabular}{|c|c|}
 \hline
 1 & 2\\
 \hline
 model& predictive RMSE\\
 \hline
 1&  0.8439727    \\
 2&  0.8417796    \\
 3&  0.8401334    \\
 4&  0.8433204    \\
 5&  0.8456246    \\
 6&  0.8500992    \\
 7&  0.8541917    \\

 \hline
\end{tabular}
\caption{ The predictive RMSE scores for the different
models.}\label{it02}
\end{table}

\newpage
\begin{figure}[!hb]
  \begin{center}
    \scalebox{0.8}{\includegraphics{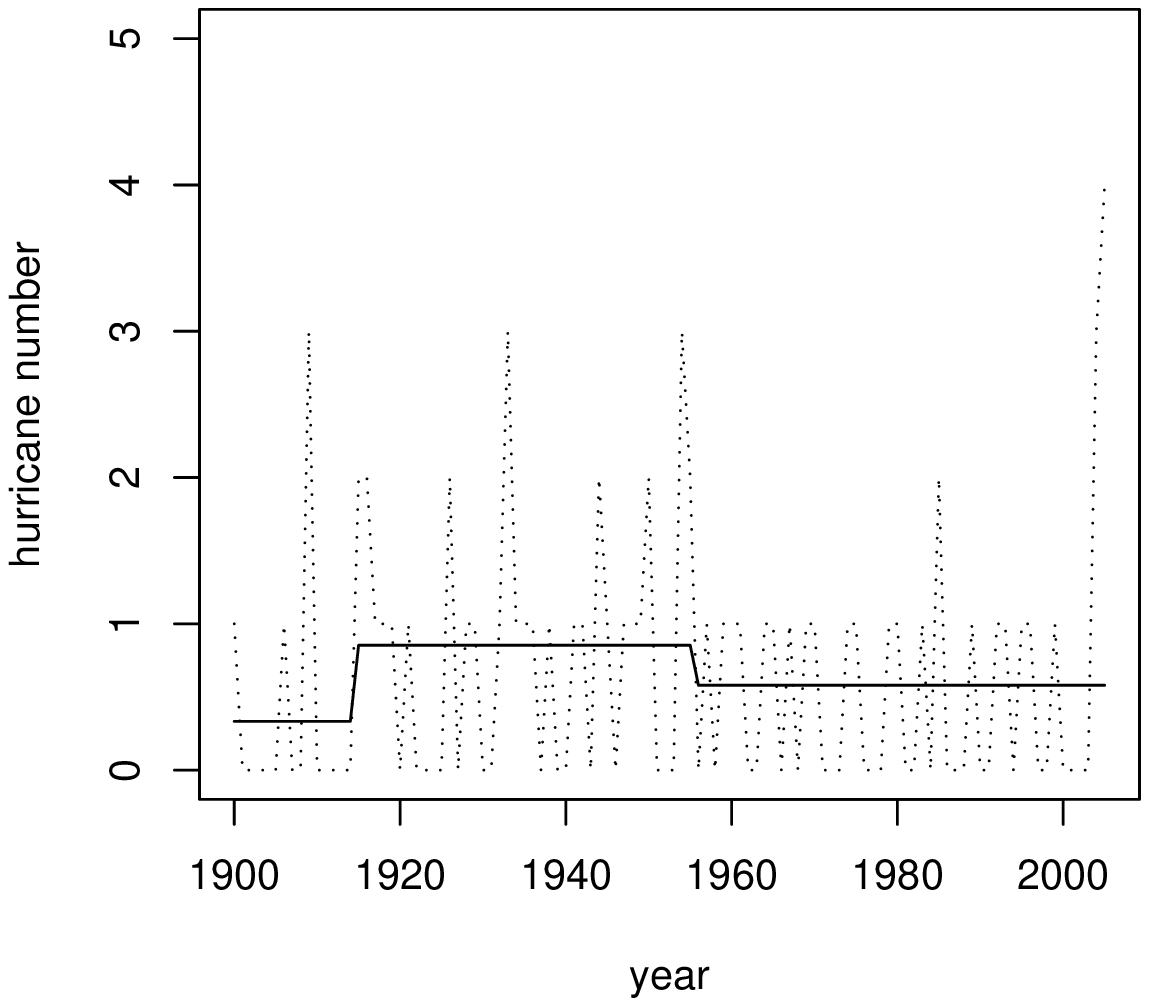}}
  \end{center}
    \caption{
The best 3 level model. }
     \label{if03}
  \begin{center}
    \scalebox{0.8}{\includegraphics{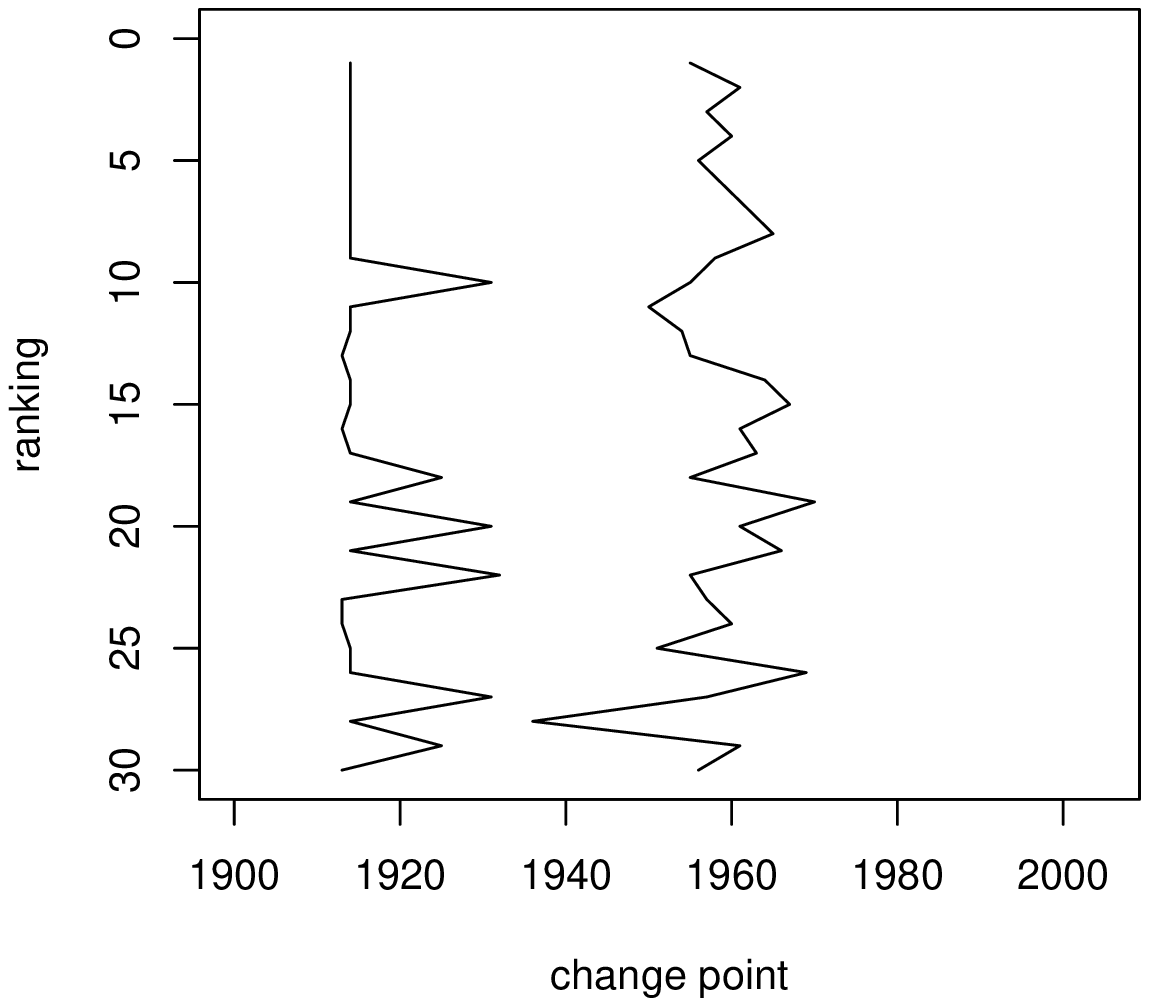}}
  \end{center}
    \caption{
The change points for the top 30 three level models. }
     \label{if04}
\end{figure}

\end{document}